\documentclass[review,number,sort&compress]{elsarticle}
\usepackage[utf8]{inputenc}
\usepackage{graphics}
\usepackage{epsfig}
\usepackage{textcomp}
\usepackage{amssymb}
\usepackage{amsmath}
\usepackage{mathptmx}
\usepackage{graphicx}
\usepackage{amssymb}
\usepackage{epsfig}
\usepackage{lineno}
\usepackage{xcolor}
\usepackage[normalem]{ulem}

\biboptions{comma,square,numbers,sort&compress}

\journal{NIM A}

\begin{document}

\begin{frontmatter}

\title{Enhanced Tensor Polarization in Solid-State Targets}

\author{D.~Keller}
\author{D.~Crabb}
\author{D.~Day}

\address{University of Virginia Physics Department, Charlottesville, Virginia 22901}

\begin{abstract}
We report measurements of enhanced tensor polarization on solid-state targets.  The results here represent an increase in tensor polarization over that previously achieved in high energy and nuclear scattering experiments that
focused on the measurement of tensor polarized observables.  Enhancement techniques are used which require RF produced close to the Larmor frequency of the target spins and use selective semi-saturation resulting from two sources of irradiation, microwave for the DNP process and the additional RF used to manipulate the population of the energy levels in the target material.  The spin dynamics of the solid target are used to align the spins enhancing the ensemble average to improve the figure of merit of the scattering experiment.  Target rotation at an optimized rate can lead to additional enhancement by applying selective semi-saturation in polycrystalline materials that possess a Pake doublet in their NMR signal.
\end{abstract}

\end{frontmatter}

\section{Introduction}
\label{intro}
Dynamic nuclear polarization (DNP) of targets using solid diamagnetic material, doped with paramagnetic radicals, provides a considerable advantage
in interaction rate over gaseous targets when limited to ($\sim6\times 10^{11}$ e/sec) equivalent beam intensity.  For a 1 K and 5 T system, dynamically enhanced targets can be used in vector polarized proton and deuteron experiments near 98\% and 50\% respectively for this beam intensity range.
These high cooling power DNP target systems have been used to reach luminosities up to $10^{36}$ cm$^{-2}$ s$^{-1}$.  These systems contain a high powered microwave generator, a superconducting magnet ($\sim$5T) to produce the Zeeman splitting, and a high cooling power evaporation refrigerator used to hold the target at 1 K despite the
heat-load from the beam and microwaves \cite{crabb1}. 

Polarization of spin-1 deuterons can be achieved using DNP with deuterated materials like ND$_3$, C$_4$D$_9$OH, or LiD.  In the spin-1 target, the deuterons have nonzero quadrupole moments, and the structural arrangement of the nuclei in the solid generate electric field gradients (EFG) which couple to the quadrupole moment.  This results in an additional degree of freedom in polarization that the spin-1/2 nucleons do not possess.  The target spins in the ensemble can be aligned in both a vector ($P$) and a tensor ($P_{zz}$) polarization.  Defined
in terms of the relative occupation of the three magnetic substates of the spin-1 system ($m=0$, $\pm1$), they are,
$$P=\frac{n_+ - n_-}{N}$$ and
$$P_{zz}=\frac{n_+ -2n_0 + n_-}{N},$$
where $n_+$, $n_0$, and $n_-$ being are the relative occupation of the magnetic substate $m$, and the total population is $N=n_+ + n_0 + n_-$.

Interest in tensor-polarized observables \cite{kumano1,kumano2,kumano3,kumano4,kum1,kum2,jaffe,Day,Duke,JP} has prompted a new research effort to develop a maximally enhanced solid-state tensor polarized target that can be used under continuous beam application.  These observables are unique, allowing access to information not available using vector-polarized targets alone. There is still very little scattering data on tensor-polarized observables from solid polarized target experiments to date \cite{smith1,smith2,otter1,smith3,otter2}.  This is because it is necessary to maximize both the tensor polarization and beam intensity to achieve sufficiently high precision.  This requires a target system that can withstand the heat load of a high intensity beam,
but also can produce and hold high tensor polarization.

This article reports on a set of experimental measurements of enhanced tensor polarization for a solid-state target in a 1 K and 5 T system using the method of selective semi-saturation.  The details of the technique and measurement theory have been discussed previously \cite{kel1}.
This article is organized as follows.  In Section 2, a description of the spin-1 NMR lineshape is given, along with the definitions of vector and tensor polarizations and how they relate to the lineshape of the materials of interest.  The tensor enhancement technique is discussed in Section 3.  Section 4 provides the experimental techniques used to obtain the NMR measurements. Section 5 contains the results for two methods. The first method uses semi-saturating RF enhancement and other the second uses this same technique while rotating the target cell.  There is also a brief discussion on other RF techniques in this section.  Concluding remakes are given in Section 6.

\section{The Deuteron NMR Lineshape}
\label{signal}
The quadrupole moment of the spin-1 nuclei results from the nonspherically symmetric charge distribution in the quadrupolar nucleus.  For materials without cubic symmetry (e.g. C$_4$D$_9$OH or ND$_3$), the interaction of the quadrupole moment with the EFG breaks the degeneracy of the energy transitions, leading to two overlapping absorption lines in the NMR spectra. 
\begin{figure}
\begin{center}
\includegraphics[height=67mm, angle=0]{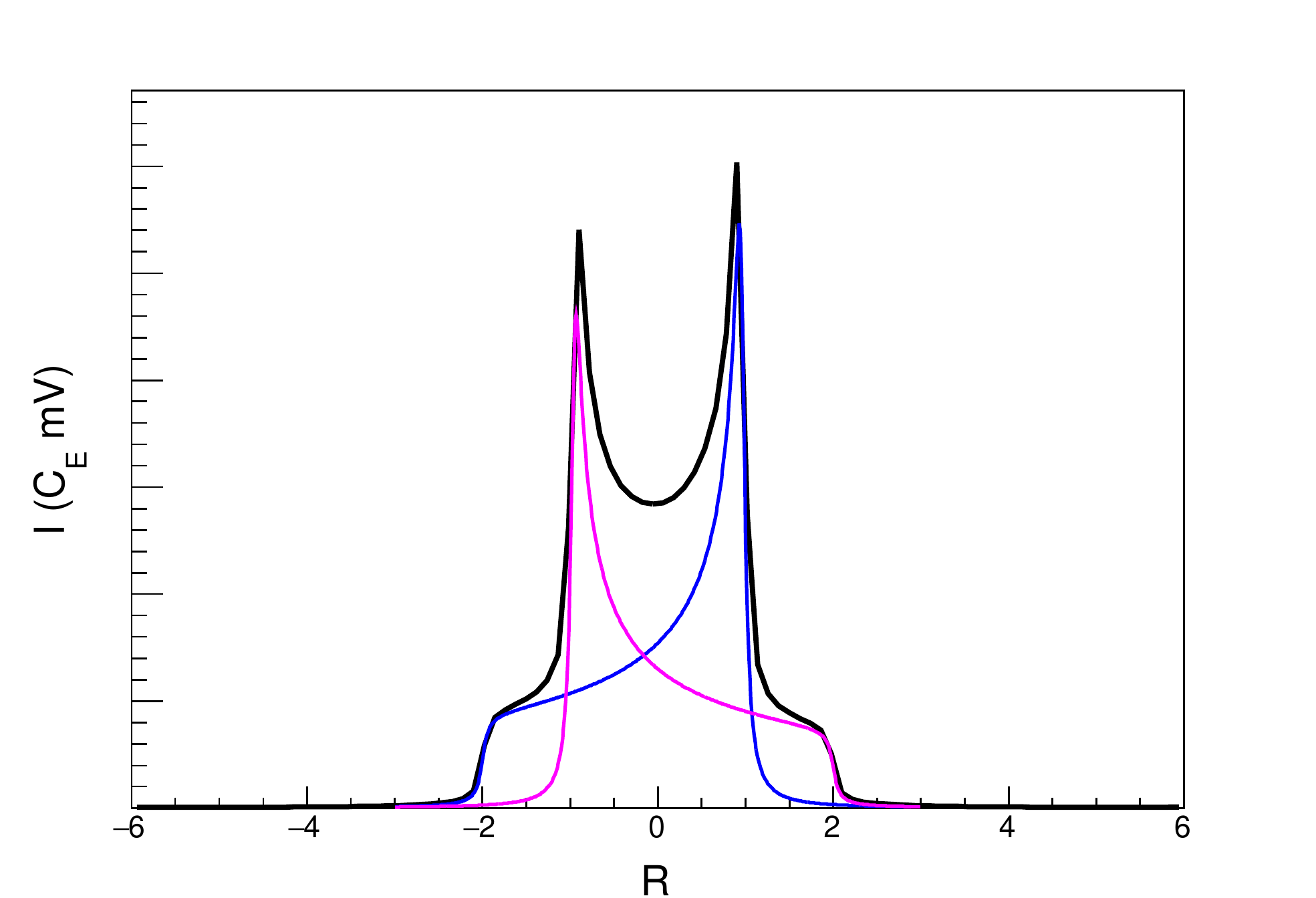}
\end{center}
\caption{An example of the NMR lineshape of a spin-1 target with a non-cubic symmetry demonstrating the two overlapping absorption lines.  The two intensities of the signal $I_{+}$ and $I_{-}$ are shown in blue and pink respectively. }
\label{pake}
\end{figure}
The spin-1 NMR lineshape is shown in Fig. \ref{pake}, demonstrating the two intensities $I_+$ (in blue) and $I_-$ (in pink).  In terms of population,$$I_+=C(\rho_+-\rho_0)$$ and $$I_-=C(\rho_0-\rho_-),$$ where $\rho_x$ is the population density in the $m=x$ energy level, and
$C$ is the calibration constant.
The term intensity is used here to indicate both the height and area of these two individual regions.  The frequency is indicated by a dimensionless position in the NMR line $R=(\omega-\omega_D)/3\omega_Q$ which spans the domain of the NMR signal, where $\omega_Q$ is the quadrupolar coupling constant.  In these units, $R=0$ corresponds to the Larmor frequency of the deuteron at 5 T ($\omega_D=32.679$ MHz).
The local electric field gradients that couple to the quadrupole moments of the spin-1 system cause an asymmetric splitting of the energy levels into two overlapping absorption lines.  The energy levels of
the non-cubic symmetry spin-1 system can be expressed
as,
$$E_m=-\hbar \omega_D m + \hbar \omega_Q (3\text{cos}^2\theta-1+\eta \text{sin}^2\theta \text{cos}2\phi)(3m^2-2),$$
where $\theta$ is the polar angle between the axis of the deuteron bond and the magnetic field, see Fig. \ref{levels}.  The azimuthal angle $\phi$ and parameter $\eta$ are fixed parameters used to characterize the electric field gradient with respect to the deuteron bond axis.
The degree of axial symmetry and dependence on the polar angle can be understood from the basis lineshape for an isotropic rigid solid which is known as a Pake doublet \cite{pake}.  The polarization information can be extracted from a fit of the NMR data providing the areas of the two intensities \cite{dulya,kel1}.
\begin{figure}
\begin{center}
\includegraphics[height=67mm, angle=0]{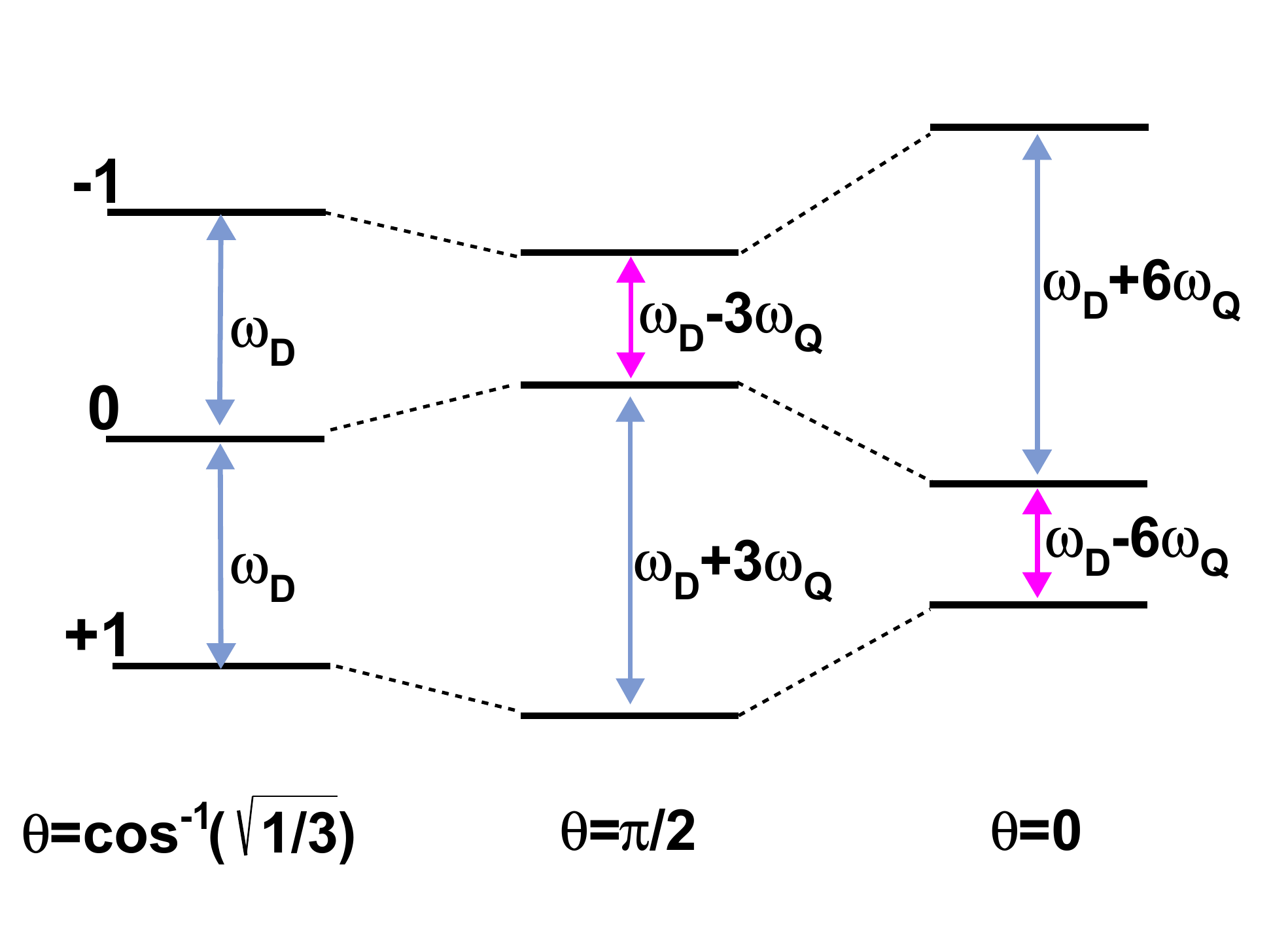}
\end{center}
\caption{The energy level diagram for deuterons in a magnetic field for three values of $\theta$ where $\hbar \omega_D$ is the deuteron Zeeman energy, and $\hbar \omega_Q$ is the quadrupole energy.  The color indicates which transition corresponds to which peak shown in Fig. \ref{pake}.}
\label{levels}
\end{figure}
The peaks of the Pake doublet ($R\sim\pm1$) correspond to the principal axis of the coupling interaction being perpendicular ($\theta=\pi/2$) to the magnetic field. This is the most probable configuration within each transition, as indicated by the height in the intensity of each peak. The opposing end in each absorption line, called the pedestal, corresponds to the configuration when the principal axis of the coupling interaction is parallel ($\theta=0$) to the magnetic field, which has much less statistical significance as indicated by the small relative height in the intensities in each transition around ($R\sim\mp2$).

If the ensemble of the spin system is in thermodynamic equilibrium, the ratio of the intensities ($r=I_{+}/I_{-}$) can be used to extract the polarizations directly \cite{dulya}.
\begin{equation}
P=\frac{r^2-1}{r^2+r+1}~~~~~~P_{zz}=\frac{r^2-2r+1}{r^2+r+1}
\label{one}
\end{equation}
or simply,
\begin{equation}
\frac{P_{zz}}{P}=\frac{r-1}{r+1}.
\label{two}
\end{equation}
The extracted information from the fit also gives the sum of the two intensities which provides the vector polarization $P=C(I_{+}+I_{-})$, while the difference provides the tensor polarization $P_{zz}=C(I_+-I_-)$. It is important to note that these two expressions remain true even if the system is not in thermodynamic equilibrium, unlike Eq. \ref{one} and \ref{two}.  Once the calibration constant $C$ is measured, these expressions can be used to extract the averaged polarizations of the ensemble over the course of the HEP/Nuclear scattering experiment \cite{kel2}.

\section{Tensor Polarization Enhancement}
\label{approach}

To manipulate the magnitude of tensor polarization during DNP pumping, a separate source of coil generated RF irradiation is used to selectively saturate some
portion of the deuteron NMR line. By applying RF irradiation at a frequency or over a frequency range (hole burning \cite{jeffries, del}), transitions are induced between the
magnetic sublevels within the frequency domain of the applied RF.  A spin-diffusion rate that is
small compared to the effective nuclear relaxation rate allows for significant changes to the NMR line via the RF, which can be strategically applied
to manipulate the spin-1 tensor polarization.  In the presented set of measurements, DNP microwaves were used, as well as an additional RF source that used semi-saturating RF (ss-RF) irradiation to maximize the tensor polarization for the a 1 K and 5 T system \cite{kel1}. A semi-saturated steady-state condition was used which
manipulated and held the magnetic sublevels responsible for polarization enhancement. The continuous wave NMR (CW-NMR) lineshape was measured and manipulated to maximize tensor polarization.  The technique of ss-RF requires using a power profile that is sensitive to the intensity distributions with the correct modulation time signature over the frequency domain to optimally enhance. To be useful in a scattering experiment setting, the target
ensemble averaged tensor polarization must be increased and held in a continuous mode for application at continuous beam facilities such as Jefferson Lab.  Temporarily enhanced states during beam-target interactions are also possible for facilities that have a short beam spill per cycle such as Fermilab.

The source of ss-RF comes from a dedicated coil
with a field $\bold{B}_{\nu}$ resulting in an induced transition rate proportional to \cite{kel1},
\begin{equation}
\xi=2\pi\frac{\bold{B}_{\nu}a_{\nu}}{\bold{B}_0}\delta(\omega_D-\omega_{\nu}),
\end{equation}
where $\omega_D$ is the Larmor frequency, $\omega_{\nu}$ is the ss-RF frequency, $a_{\nu}$ is the coupling constant, and $\bold{B}_0$ is the strength of the holding field.  The ss-RF can only play a role between nuclear spin energy levels that differ by the spin of the mediating photon.  The ss-RF drives transitions that lead to equalization of the populations in the energy levels at the applied frequencies.  This implies that the change in intensities at the location $R$ in the NMR line due to the ss-RF can be expressed as,
\begin{equation}
\frac{I_{\pm}(R)}{dt}=-2\xi\omega_1 I_{\pm}(R),
\label{dos1}
\end{equation}
\begin{equation}
\frac{I_{\mp}(-R)}{dt}=\xi\omega_1 I_{\pm}(R).
\label{dos2}
\end{equation}
In other words, the rate at which any one of the intensities change due to ss-RF is only dependent on
the intensity level and the strength of the $\bold{B}_{\nu}$ field, or RF power.  Here $\omega_1$ is the reciprocal of the electron longitudinal relaxation rate used to be consistent with previous work \cite{kel1}.  The total polarization can only be decreased at the ss-RF location in $R$, so strategic implementation is required to enhance the difference in the integrated $I_+$ and $I_-$ regions.  Equation \ref{dos1} describes the rate of change in the intensity driven by the ss-RF for any region at $R$.  The transitions driven by the ss-RF decrease the intensity at $R$.  However, the coupling to Equation \ref{dos2} indicates an increase in intensity in the opposing absorption line at $-R$ at a rate that is half as fast as the decrease in intensity at $R$.  This also implies that the region at $-R$ increases in area by half of the lost area at $R$.

Equations \ref{dos1} and \ref{dos2} affirm that materials with the same lineshape can be treated exactly the same under ss-RF tensor enhancement.  This expression does not change for different materials relaxation rates.  For many of our experimental results, we use deuterated butanol (C$_4$D$_9$OD) (99\% deuterated) rather than deuterated ammonia (ND$_3$) only because the paramagnetic complex of deuterated butanol is easier to optimize, but the maximum total polarizations and lineshapes are very similar.

\section{Experimental Techniques}
\begin{figure}
\begin{center}
    \includegraphics[width=0.60\textwidth]{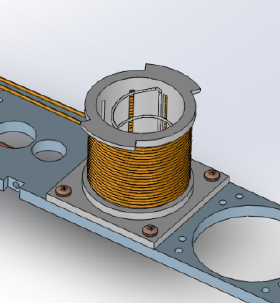}
\end{center}
\caption{Drawing of the ss-RF cup and coil used for the set of experiments discussed.  The NMR coil is also shown inside.}
\label{cup}
\end{figure}
All experimental results were taken at the University of Virginia's Solid Polarized Target Lab. The experimental data was taken
at 5 T and 1 K with an evaporation refrigerator \cite{crabb1,crabb2} that had a cooling power of just over 1 W, and a DNP microwave source generating approximately 0.3 W on the target.  The nuclear spin polarization was measured with an CW-NMR coil and Liverpool Q-meter \cite{court}. With this system, the RF
susceptibility of the material was inductively coupled to the NMR coil, which was part of a series LCR Q-meter circuit that was tuned to the Larmor frequency of the nuclei of interest.
The Q-meter based NMR provided a non-destructive polarization probe of the nuclear spin ensemble in the solid-state target.  

For the selective excitation using ss-RF, an additional coil around the target cup was connected to an RF-generator. The ss-RF coil consisted of 20 turns of silver covered copper clad non-magnetic steel with a diameter of $\sim$0.2 mm.  The coil was constructed to approximate a homogeneous RF field around the target material that was perpendicular to the holding field.  For optimal performance, the coil was both impedance matched and tuned as an LCR circuit to maximize power and reduce reflection at the coil.  The target cell used was 2.5 cm in diameter and 3 cm long with a volume of about 15 cm$^3$, which held about 9 g of d-butanol.  The ss-RF coil was fitted and fixed around the diameter of the target cell.  The ss-RF was modulated over the frequency domain of interest at the appropriate RF power to semi-saturate the NMR line to the intensity level of interest.

Adequate power delivery to the material from the ss-RF source was critical.  Even with the ss-RF coil matched and tuned, amplification was required as the SMT03 Rohde \& Schwarz (R\&S) RF generator used only delivered a maximum of 20 mW.  We used an inline RF amplifier from Minicircuits (model LZY-1+) at 50 $\Omega$ and with 50 W amplification between 20 to 512 MHz.  The power requirements at the coil was on the order of 10$^2$ mW.

The software controlled power specifications and RF modulation were handled in LabView.  The computer controlled
R\&S RF Generator could be made to change
amplitude and duration at each frequency step, enabling
a highly selective degree of spin manipulation as a function of time.  In this initial study, we did not optimize this feature.
The distribution of the power profile delivered was also dependant on the coil.  A high quality factor and optimized tune in the ss-RF coil delivered a more precise and localized RF load in the signal domain.  The power profile of the ss-RF in the CW-NMR was characterized by the Voigt function \cite{kel1}.

The target material used in these measurements was
warm irradiated ($\sim$87 K) under liquid argon using 10 MeV electrons, with integrated fluxes of about $10^{17}$ e$^{-}$/cm$^{2}$ at the MIRF facility at NIST.
The production of the paramagnetic complex in irradiated
C$_4$D$_9$OH conditioned the sample to polarize to about $P=50\%$ at 5 T and 1 K.
Under Boltzmann equilibrium (no ss-RF), this resulted in a tensor polarization of,
\begin{equation}
P_{zz}=2-\sqrt{4-3P^2}=19.7\%.
\label{boltz}
\end{equation}
Enhancement beyond this level requires application of selective excitation using the ss-RF to maximize the difference in the two intensities $I_+$ and $I_-$ such
that $P_{zz}=C(I_{+}-I_{-})$ is maximized.

\begin{figure}[!tbp]
  \centering
  \begin{minipage}[b]{0.45\textwidth}
    \includegraphics[width=\textwidth]{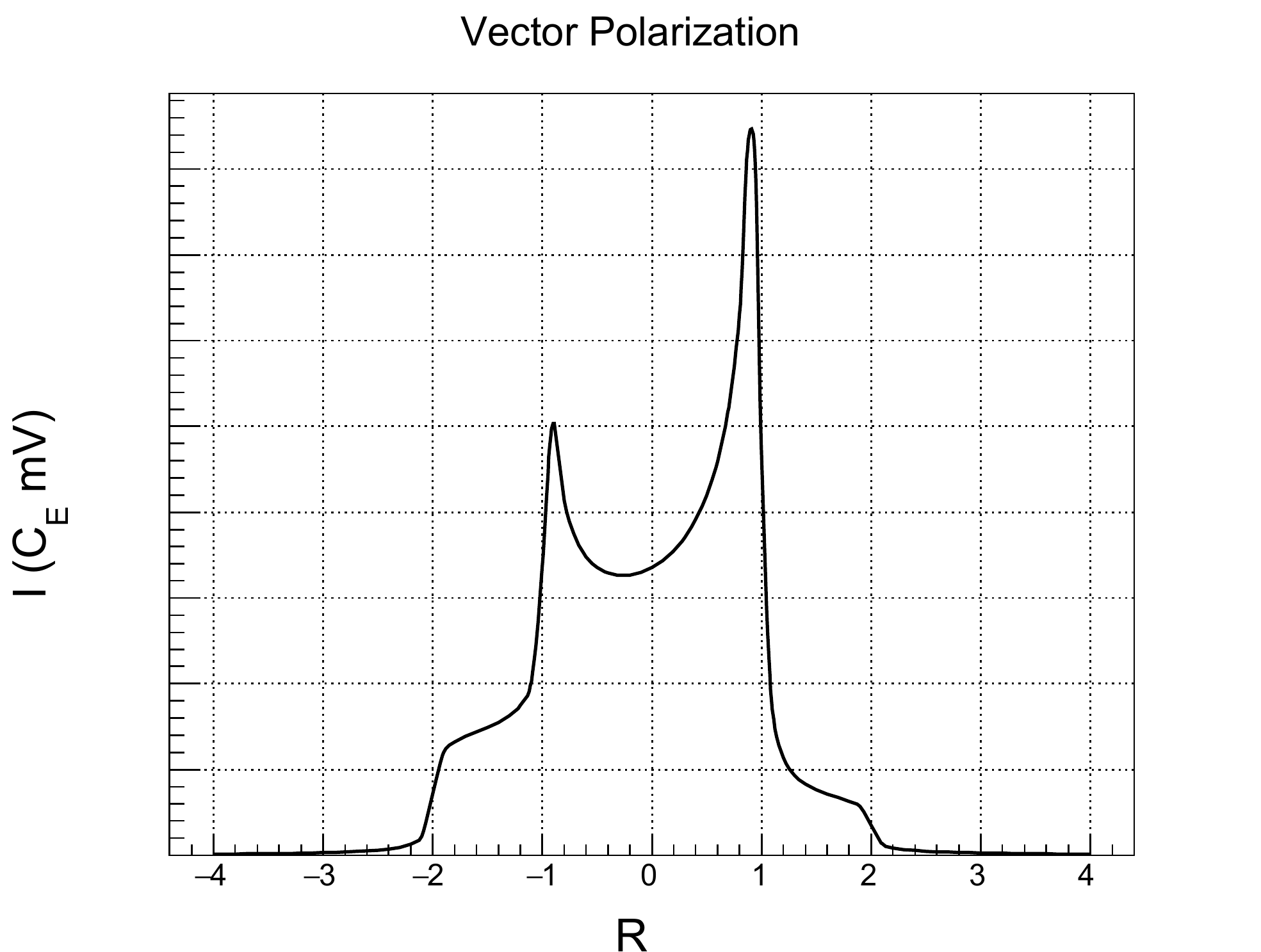}
    \caption{The NMR lineshape for a vector polarization
    of 50\% which comes from the sum of the intensities $I_+(R)$ and $I_-(R)$.}
    \label{vec}
  \end{minipage}
  \hfill
  \begin{minipage}[b]{0.45\textwidth}
    \includegraphics[width=\textwidth]{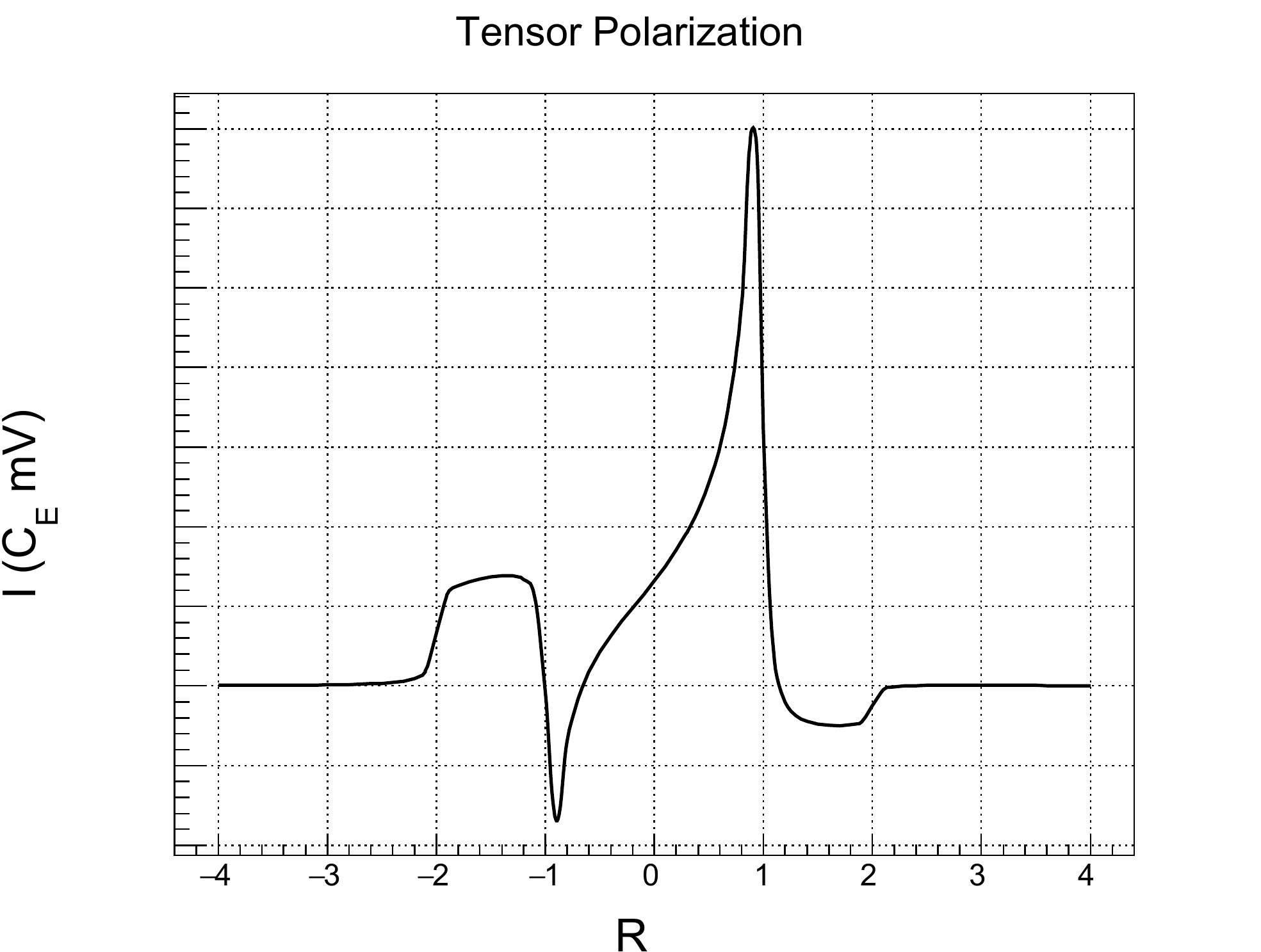}
    \caption{The tensor polarization shown from the
    difference of the intensities $I_+(R)$ and $I_-(R)$.}
    \label{tens}
  \end{minipage}
\end{figure}

Figure \ref{vec} shows a plot of the NMR lineshape for
a vector polarization of 50\%.  This plot represents the sum of $I_{+}(R)$ and $I_{-}$(R) over the frequency domain in $R$.  Similarly, a tensor polarization plot is shown in 
Figure \ref{tens} and represents the difference of $I_{+}(R)$ and $I_{-}(R)$ over the frequency domain in $R$.  By selectively applying the ss-RF, it was possible to reduce the regions in the $P_{zz}$ line that drop below the x-axis.  When this was done simultaneously over all negative regions in the domain, the tensor polarization was enhanced.

\section{Experimental Results}
\subsection{Semi-Saturating RF Enhancement}
To optimize the enhancement, the ss-RF excitation had to minimize the negative tensor polarization for all $R$ while minimizing the reduction to the overall area of the NMR signal from the process.  The two critical regions lie around $R\sim\mp1$ ($\theta\approx\pi/2$) and $\pm1<R<\pm2$  ($\theta \approx 0$). For positive vector polarization, as in Fig. \ref{vec}, the greatest integrated tensor polarization enhancement was achieved through selective excitation to reduce the size of the smaller transition area with intensity $I_{-}$.  This can be thought of as minimizing the negative parts of the tensor polarization, shown in Fig. \ref{tens}.  In both figures, the y-axis would normally be millivolts scaled by a multiplicative factor $C_E$ which is sensitive to the characteristics of the NMR coil, such as inductance, geometry, and orientation.  We therefore leave these units generalized and provide a scale for relative change when necessary.  For negative vector polarization, the greatest enhancement came from the reduction of the transition area with intensity $I_{+}$. Otherwise, the treatment of both cases was identical, so it was convenient to focus on positive vector and tensor polarization.

The target had to first be polarized with DNP to achieve the highest vector polarization possible for that material.  This maximized the signal area to be used in the ss-RF manipulation.  For C$_4$D$_9$OH, the polarization started to maximize
at about 50\% in about 35 minutes.  For ND$_3$, the time to polarize would be significantly longer, but it can reach the same maximum.

The ss-RF was then applied as described.  Because of the power amplification in the ss-RF, damage to the Q-meter may result if these systems are running simultaneously.  Cycling between RF manipulation and NMR measurement can result in additional uncertainty in the NMR measurement due to the delayed sampling and the evolution of the spin state in the ensemble over time.  In the results presented, the Q-meter and ss-RF were run simultaneously near the end of the ss-RF cycle to acquire the best NMR measurements before the manipulated signal relaxed back to Boltzmann equilibrium.  This still required considerable synchronicity between these two RF systems to make the most of the limited NMR sweeps after the ss-RF was turned off and before the signal lineshape changed from relaxation.  The Q-meter NMR sampling rate was set to about 15.4 Hz over the 500 steps in the domain where each bin represented one of the 500 RF scan steps of the CW-NMR.  For the sake of accurate measurement of the ss-RF enhancement to the signal, the material spin diffusion relaxation rate became very relevant.  This relaxation rate was reduced as much as possible by maintaining low temperature ($\sim$1 K), turning off the DNP microwaves, and starting the NMR initial sweep as close as possible to the time that the ss-RF stopped.

The ss-RF was applied by modulating the frequency over the domain of interest. The pedestal region of the smaller intensity ($I_-$) could be brought to near saturation to optimize.  Saturation occurred when the RF drove the population of the magnetic sublevels to equalize.  However, the peak was highly sensitive to power being higher in magnitude, and it was necessary to preserve as much of the larger intensity ($I_+$) underneath the $I_-$ peak as possible.  Optimization required just the right amount of RF power to reduce the area in $I_-$ without depleting $I_+$.

\begin{figure}[!tbp]
  \centering
  \begin{minipage}[b]{0.45\textwidth}
    \includegraphics[width=\textwidth]{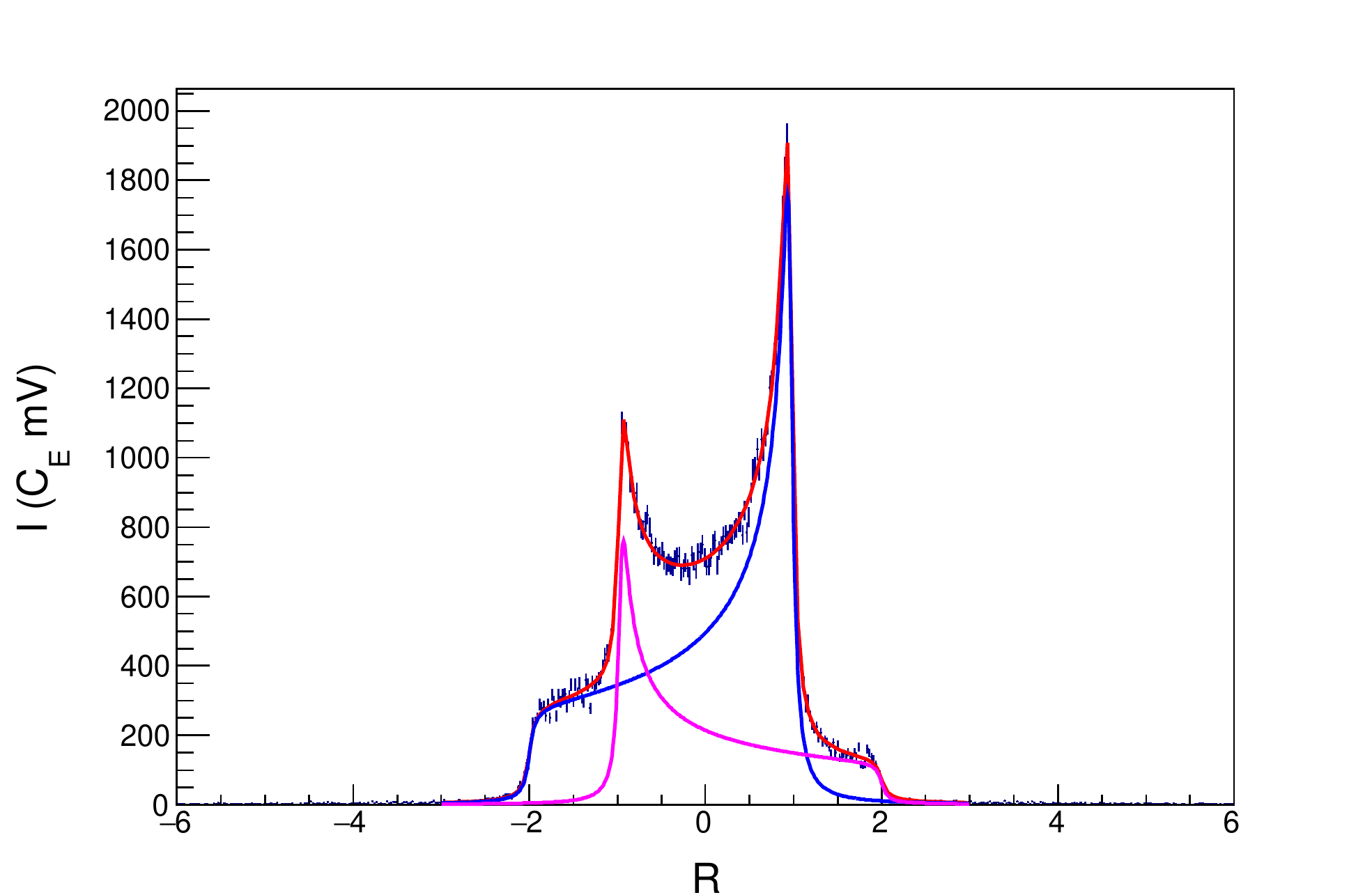}
    \caption{The NMR measurement for a deuteron polarization
    at 49.8\%, with the fit used to extract the polarization measurement.}
    \label{dat1}
  \end{minipage}
  \hfill
  \begin{minipage}[b]{0.45\textwidth}
    \includegraphics[width=\textwidth]{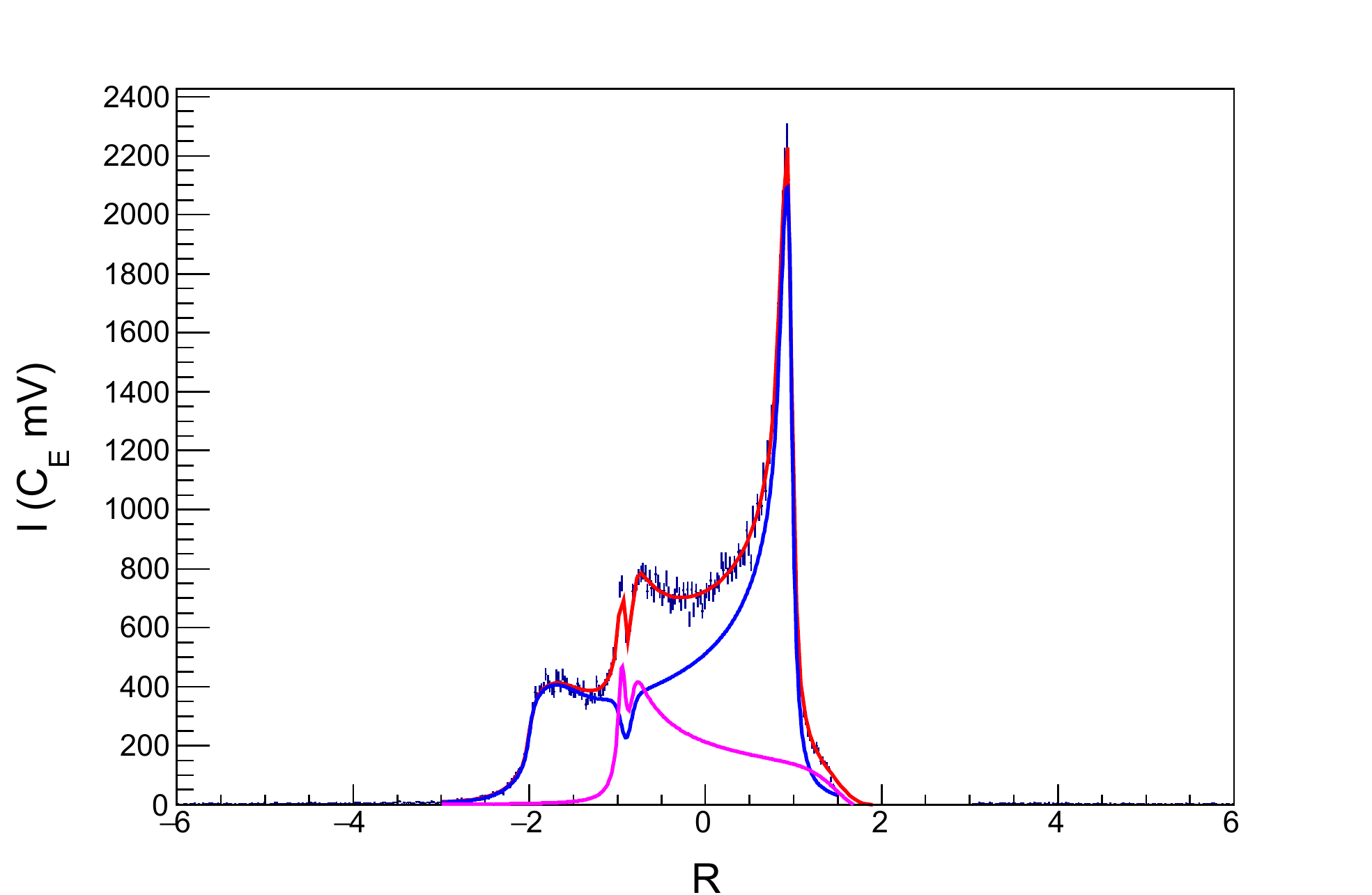}
    \caption{NMR measurement with fit result after the ss-RF has been applied to the two negative tensor regions.  The tensor
    polarization from this example was 28.8\%.}
    \label{dat2}
  \end{minipage}
\end{figure}

Figure \ref{dat1} shows the NMR measurement right before ss-RF manipulation.  The vector polarization was 49.8\%.  A fit to the data is also shown indicating the $I_+$ intensity in blue and the $I_-$ intensity in pink.  Figure \ref{dat2} shows the NMR measurement after the ss-RF had been applied to the two negative tensor regions in $R$.  A fit to the data is shown which uses only the constraints from Eq. \ref{dos1} and Eq. \ref{dos2}, resulting in a tensor polarization measurement of 28.8\%.  Several iterations were made to optimize enhancement given variation in the set of RF parameters.  After several trials, it was determined that the best RF waveform was triangular with modulation at around 1 kHz.  The duration of the step where the modulation was applied could also be a critical parameter, because the saturation level under DNP is sensitive to both power and time of modulation.  In the following, we held the duration constant across the applied modulation to reduce the number of variables.  We then waited for the system to reach a steady-state and then recorded the results.  We present the parameters for the best measurements with the cleanest signal and best fit results in Table \ref{table:1}.

\begin{table}[h!]
\begin{center}
\begin{tabular}{ |p{2.2cm}|p{1.1cm}|p{2.2cm}|p{1.1cm}||p{1.1cm}|p{1.1cm}|  }
 \hline
 \multicolumn{6}{|c|}{ss-RF Enhanced Measurements} \\
 \hline
 Peak (MHz) & Amp (mV) & Pedestal (MHz) & Amp (mV) &$P_{zz}$ (\%) & Error (\%)\\
 \hline
 32.62(0.000) & 20 & 32.85(0.015) & 70 & 26.7 &5.4\\
 32.63(0.015) & 30 & 32.85(0.020) & 40 & 28.8 &5.7\\
 32.64(0.015) & 30 & 32.84(0.025) & 40 & 29.4 &7.2\\
 32.64(0.015) & 25 & 32.83(0.035) & 20 & 26.5 &6.8\\
 32.64(0.015) & 20 & 32.85(0.035) & 70 & 30.3 &7.8\\
 32.64(0.020) & 20 & 32.85(0.025) & 40 & 27.5 &4.7\\
 32.64(0.015) & 40 & 32.88(0.055) & 50 & 31.1 &8.5\\
 \hline
\end{tabular}
\end{center}
\caption{NMR measurements for various trials of optimized tensor enhancement for different applied ss-RF amplitude and frequency location for both the pedestal and peak, and the corresponding modulation range.}
\label{table:1}
\end{table}

For the set of measurements listed in Table \ref{table:1}, the amplitude (Amp) and modulation center frequency and range are given for the RF applied to the peak and pedestal.  The resulting tensor polarization was extracted after the data was taken and analyzed offline. The listed errors were calculated from the systematic error from the fit and the standard NMR uncertainty measurements \cite{kel2}.  All the errors listed in the table are relative contributions.  For all of this data, the initial vector polarization was very close to 50\% with an
uncertainty of 3.2\% relative.  To minimize the uncertainty, multiple thermal equilibrium calibration measurements were performed at different temperatures while cross checking with the standard intensity fitting procedure \cite{kel1,dulya}.  The conclusion deduced from this set of data is that optimal enhancement comes from the selection of the appropriate frequencies and modulate range without applying any additional unnecessary power.  Ultimately, it is necessary to implement the ss-RF with a power profile that is sensitive to the lineshape over the frequency domain while minimizing the overall RF-power load to the target.

 The same techniques that are used to enhance the tensor polarization can also be used to reduce it. This is done by applying the ss-RF to the larger peak and manipulating the signal so that $I_-$ and $I_+$ are equal.  We performed tests of this application and measured approximately 37\% vector polarization with tensor polarization of 0.02 with a 3\% relative error.  The target was initially polarized to vector polarization of about 50\%.  This approach is helpful for spin-1 observable extraction where tensor polarized observables can contaminate the measurement. 

There is still further opportunity to advance this approach by using NMR and ss-RF that are optimally synchronized.  Additionally, the power profile of the ss-RF should be delivered across the domain as a function of the lineshape.  In the results presented, this was only approximately achieved by eye and not computationally.  Finer control and granularity could be achieved by automating this process to have greater ss-RF power deposition at higher intensities and less at lower intensities while monitoring the tensor polarization enhancement under continuous DNP.  Such improvements would also reduce uncertainties by making the resulting modified lineshape more smooth. Effective optimization of the enhancement through selective semi-saturation in real-time depends greatly on the continuous knowledge of the overlapping absorption lines.  

Online monitoring and enhancement would be critical during scattering experiments where  polarization decay occurs with the accumulation of dose due to changes in the paramagnetic complex.  As the polarization decays, the amount of enhancement per unit area increases so, like the microwave system, the ss-RF would have to be continuously monitored and optimized.  Hardware and software tools are under development at UVA to accommodate these needs for future experiments.

\subsection{Rotating Semi-Saturating RF Enhancement}
For a given vector polarization of 50\%, one can not achieve much over a 10\% absolute gain in tensor polarization with ss-RF alone.  This is because the smaller of the two intensities has a large portion of its absorption line completely inaccessible to the application of ss-RF.  Further, ss-RF manipulation of a broader domain would ultimately deplete the overall enhancement.

To access the spins in the overlapping region, it is possible to apply ss-RF in the non-overlapping region close to $\theta\approx0$ and rotate the target so that the manipulated portion of the line moves into the inaccessible overlapping region and a new set of spins becomes available to manipulate again at $\theta\approx0$.
If the ss-RF were to be applied continuously in the non-overlapping region while rotating at a rate just faster than the spin diffusion rate under DNP, but slow with respect to the nuclear spin relaxation rate, then additional enhancement is possible.  The enhancement condition is,
\begin{equation}
\omega_1\sigma_d(R)<\Omega(t)<\omega_1\lambda,
\label{cond}
\end{equation}
where $\Omega(t)$ is the rate of rotation at time $t$ and $\omega_1\sigma_d$ is the spin-diffusion rate, with $\sigma_d$ being the diffusion constant at the manipulated position $R$ and $\omega_1\lambda$ being the nuclear relaxation rate.

\begin{figure}[!tbp]
\begin{center} 
    \includegraphics[width=0.60\textwidth]{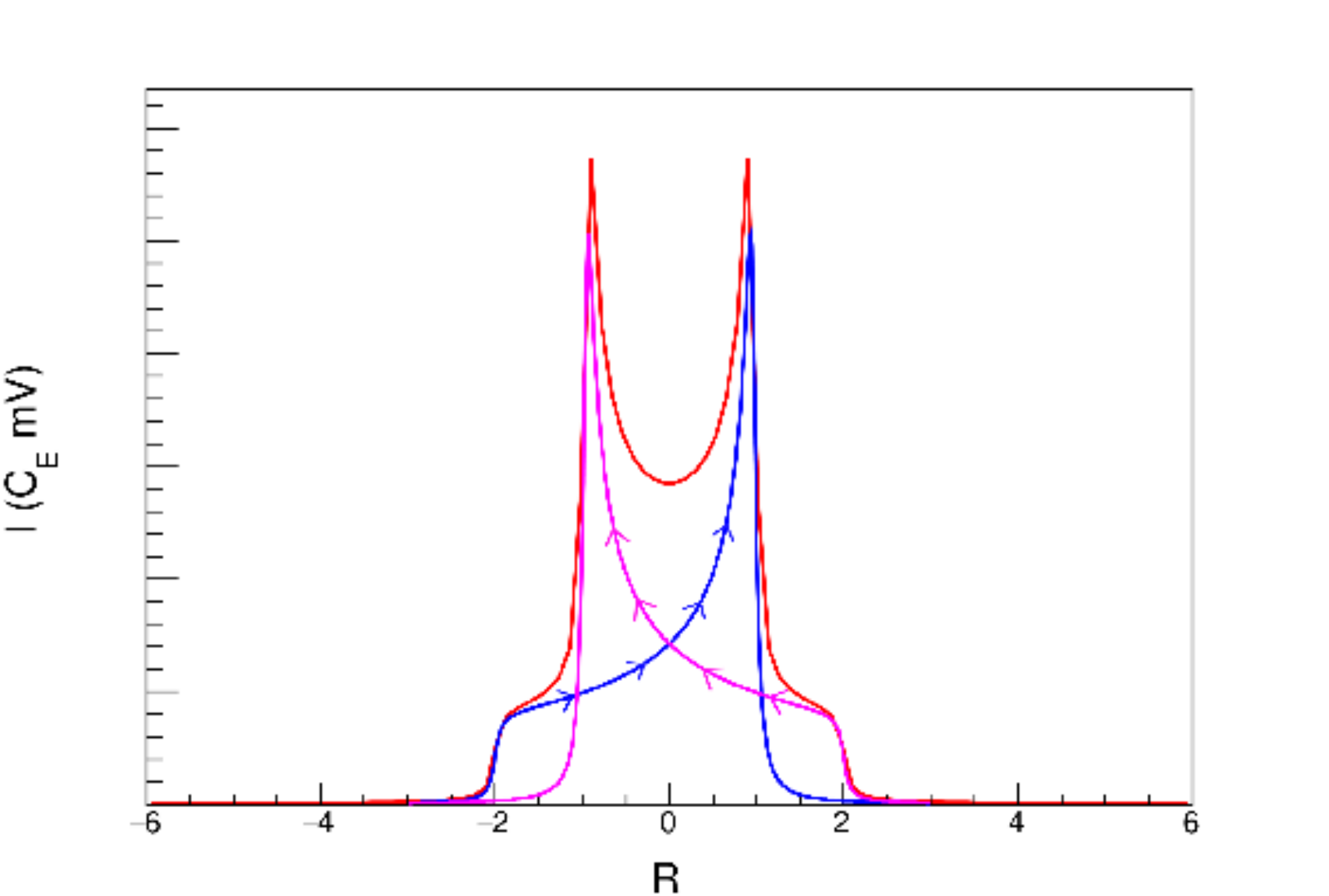}
    \caption{The direction of spins moving through the absorption lines with counter-clockwise rotation.}
    \label{rot1}
\end{center}
\end{figure}

Slow rotation of the target revolves the target crystallites over all possible orientations relative to the holding field. In this way, the local spins with any given angle $\theta$ of the internuclear vector with respect to the holding field pass through every other $\theta$ in the spectrum.  Rotation clockwise moves the spins oriented at $\theta=\pi/2$ out away to small $\theta$.  Counter-clockwise rotation moves the spins at $\theta=0$ towards $\theta=\pi/2$. Figure \ref{rot1} shows how the direction of spins move through the two absorption lines for counter-clockwise rotation.  Rotation of the target increases the accessible spins in the frequency domain providing a broader reach to the ss-RF technique.  This version of the RF technique we refer to as rotating semi-saturating RF or rss-RF.

The type of diffusion that is most critical to rss-RF is the rotation driven spectral diffusion.  This process occurs when individual nuclear spins undergo continuous exchange of energy, resulting in a transfer of polarization between spins of close proximity and neighboring frequency.  The mechanism has a $\text{cos}^2\theta$ dependence governing the fundamental rate of the flip-flop process.  Spectral diffusion increases during rotation due to the changes in spin $\theta$ orientation.  Changes occur to the line after selective excitation when nuclear spin polarization transfers across the NMR spectrum effectively refilling the depleted area to return to Boltzmann equilibrium across $\theta$.  The diffusion rate can be expressed as \cite{abr1},
\begin{equation}
\omega_1\sigma_d(\theta)\propto b^2I(\theta),
\label{cond1}
\end{equation}
\begin{equation}
b=-\frac{w}{2r^3}\left(3\text{cos}^2\theta-1\right),
\label{cond2}
\end{equation}
where $w$ is the coupling constant and $r$ is the internuclear distance between spins.  We note that Eq. \ref{cond1} is an  approximation for the target rotation orientation used here of 90$^\circ$ with respect to the holding field.  This expression is valid for slow magic angle spinning (MAS) with the axis of rotation being 54.7356$^\circ$ from the holding field and for frequencies much lower than the local field frequencies but higher than the spin-diffusion rate.

\begin{figure}
\begin{center}
    \includegraphics[width=0.60\textwidth]{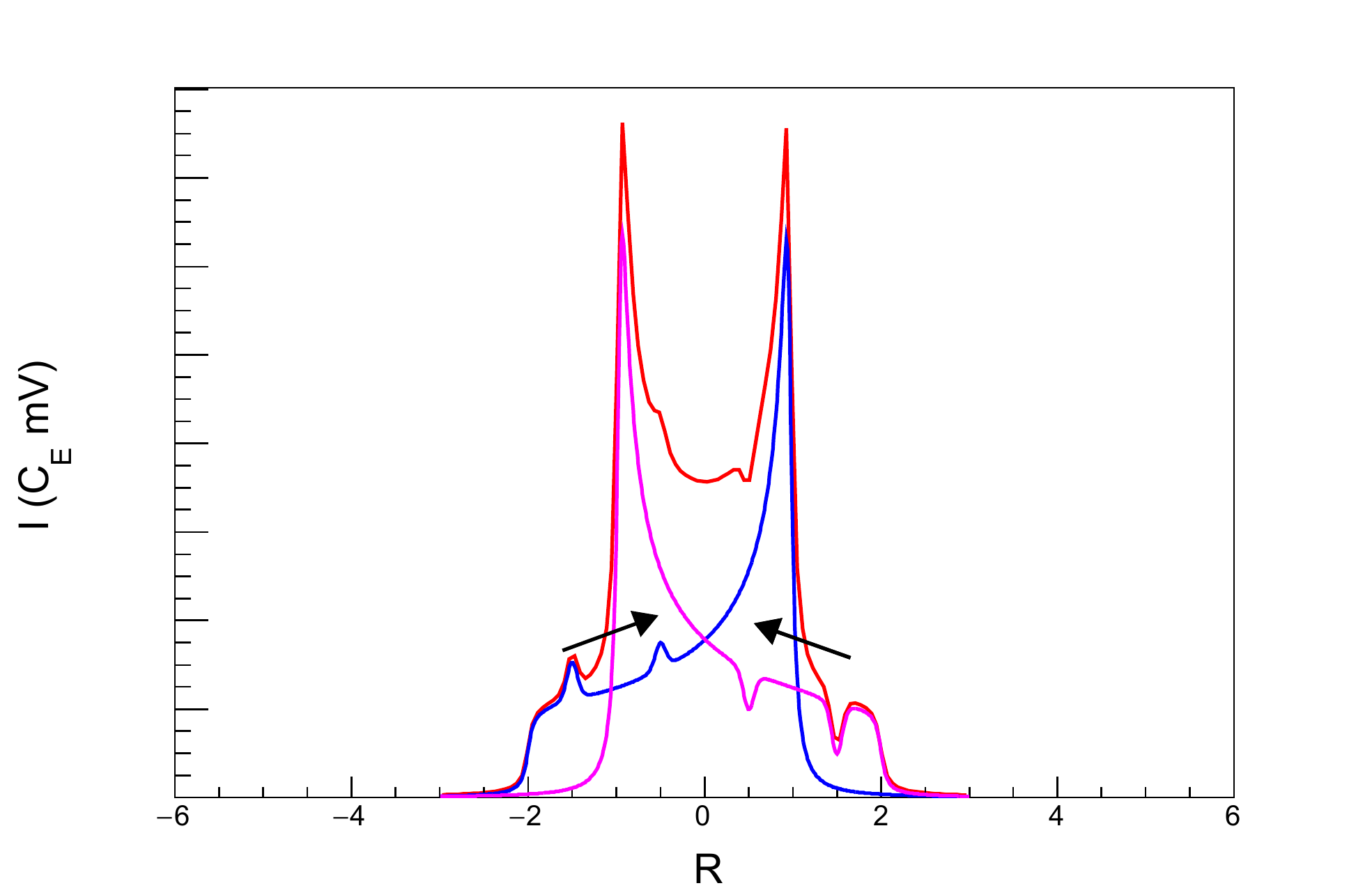}
\end{center}
\caption{A plot showing the direction of movement of a hole from ss-RF.  Both
the depleted portion and the enhanced portion in the other absorption line move
inward under counter-clockwise rotation.}
\label{rotmot}
\end{figure}

Figure \ref{rotmot} shows an example plot that illustrates the direction of movement of a hole from ss-RF.  Both the depleted portion and the enhanced portion in the opposing absorption line move
inward under counter-clockwise rotation.  Due to the rotation driven spectral diffusion, the effect is difficult to pass through the $\theta=\pi/2$ region because of the strong $\text{cos}^2\theta$ dependence.  In this regard, rotation that is too fast depletes the enhancement and can result in overall signal area reduction.

We attempted many trials of rss-RF at different rotation rates $\Omega$ and different amplitudes and
ss-RF locations in the frequency domain.  We express
the rotation rate in number of seconds in one full rotation of the target cup.  We explored a range of $\Omega^{-1}$ between 25 and 145 seconds per rotation.  Table 2 presents the best results achieved from that study, showing the resulting tensor polarization NMR measurements for various trials given the different applied rotation rate, ss-RF amplitude and frequency location for both the pedestal and peak, and the corresponding modulation range.  Again, the listed errors were calculated from the systematic error from the fit and the standard NMR uncertainty measurements.  The errors in the table are relative contributions.  For all of this data, the initial vector polarization was very close to 50\% with an
uncertainty of 3.2\% relative.  The resulting vector polarization after the rss-RF was approximately 43\%.  

\begin{table}[h!]
\begin{center}
\begin{tabular}{ |p{1.1cm}|p{2.2cm}|p{1.1cm}|p{2.2cm}|p{1.1cm}||p{1.1cm}|p{1.1cm}|  }
 \hline
 \multicolumn{7}{|c|}{rss-RF Enhanced Measurements} \\
 \hline
 $\Omega^{-1}$ & Peak (MHz) & Amp (mV) & Pedestal (MHz) & Amp (mV) &$P_{zz}$ (\%) & Error (\%)\\
 \hline
 50 & 32.65(0.010) & 15 & 32.85(0.015) & 45 & 35.7 & 8.4\\
 44 & 32.66(0.000) & 10 & 32.88(0.015) & 40 & 36.5 & 9.7\\
 40 & 32.65(0.000) & 15 & 32.88(0.015) & 40 & 36.3 & 9.3\\
 \hline
\end{tabular}
\end{center}
\caption{NMR measurements for various trials of optimized tensor enhancement for different applied rotation rate, ss-RF amplitude and frequency location for both the peak and pedestal, and the corresponding modulation range.}
\label{table:2}
\end{table}

\begin{figure}[!tbp]
  \centering
  \begin{minipage}[b]{0.45\textwidth}
    \includegraphics[width=\textwidth]{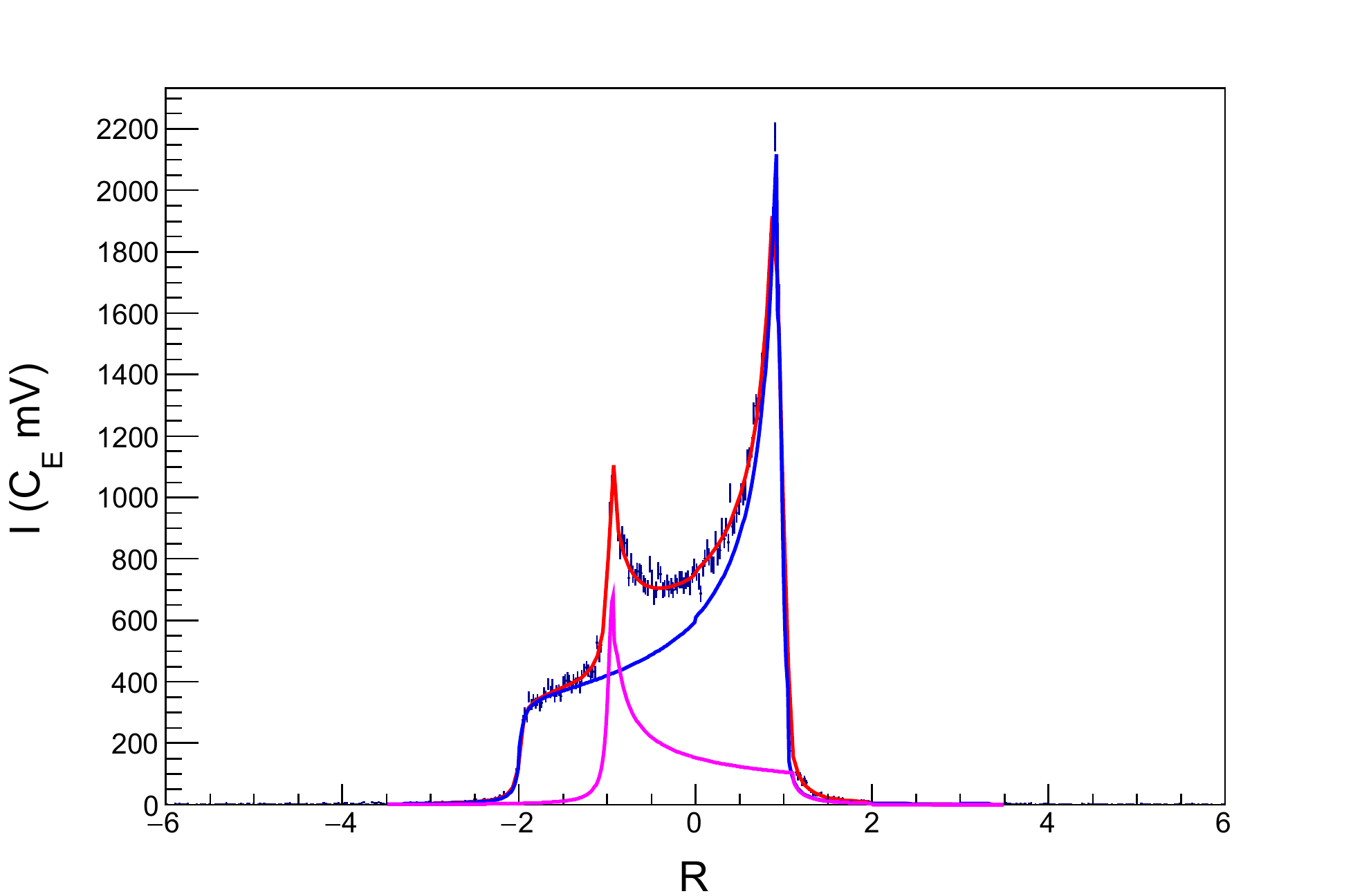}
    \caption{NMR measurement of tensor enhancement optimization that was achieved with single location of ss-RF while rotating continuously at a fixed rate.  The resulting tensor polarization was approximately 33\%.}
    \label{no1}
  \end{minipage}
  \hfill
  \begin{minipage}[b]{0.45\textwidth}
    \includegraphics[width=\textwidth]{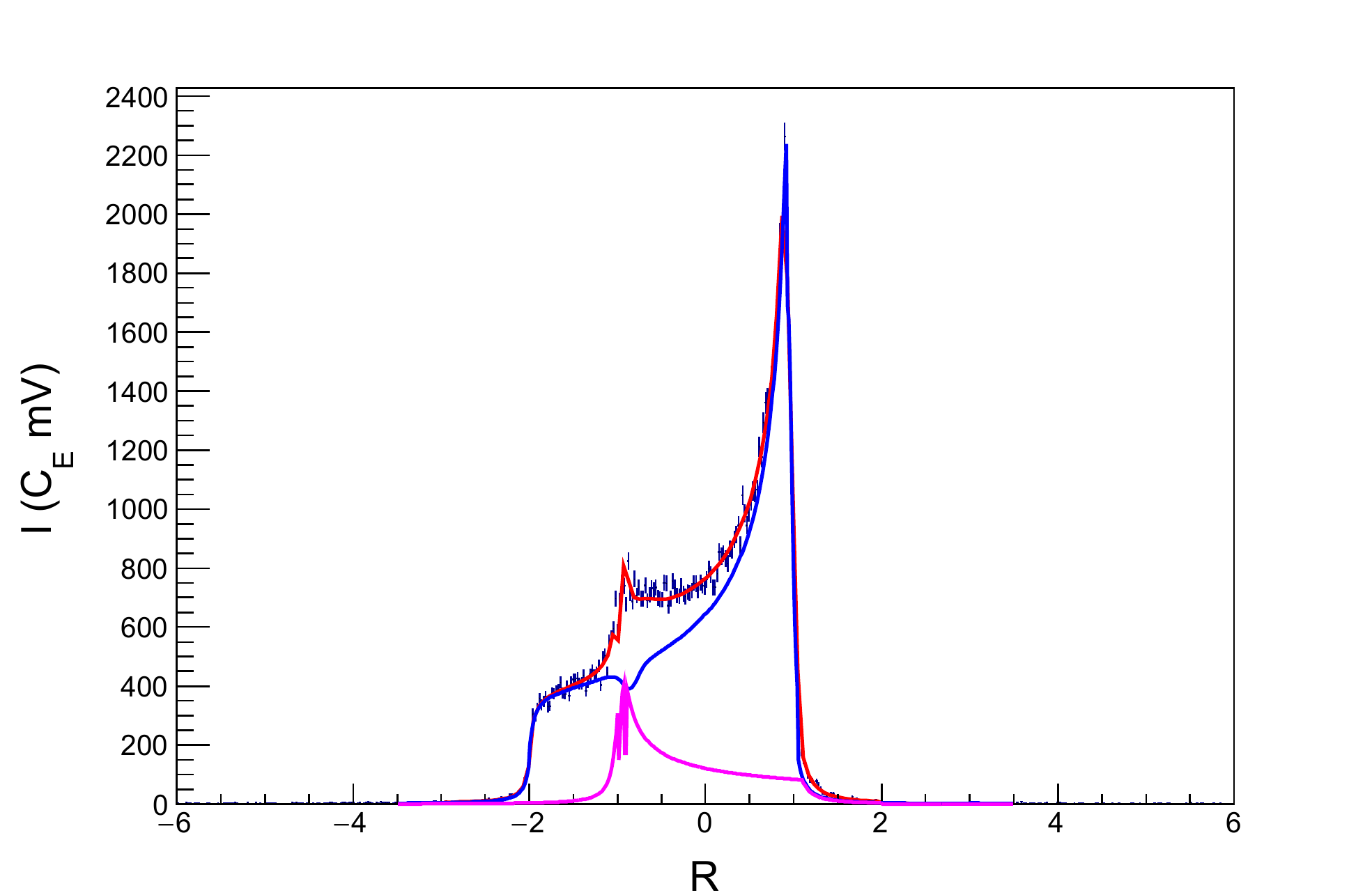}
    \caption{NMR measurement of tensor enhancement maximized by using two locations of ss-RF while rotating continuously at a fixed rate.  The resulting tensor polarization was approximately 36\%.}
    \label{no2}
  \end{minipage}
\end{figure}

With ss-RF only applied to the edge ($R\sim$2) of the smaller intensity pedestal during rotation, significant increase to the opposing absorption line could be seen, see Fig. \ref{no1}.  The opposite pedestal grew and the enhanced region moved under the smaller peak.  This is the most optimization that was achieved with this single location of ss-RF while rotating continuously over the rotation rate that was studied.  In this example, $\Omega^{-1}=40$ seconds and the resulting tensor polarization was approximately $33\%$ (8\%).

For maximal enhancement during rotation, similar application of the ss-RF in the frequency domain is required.   The RF was frequency modulated to saturate the majority of the pedestal from $R=1.5$ to $R=2$ as well as the small peak.  The location of the pedestal rss-RF was the same as it was for ss-RF, except the power was reduced as to reduce the saturation level.  The width of the RF was also reduced as to not irradiate any part of the large peak tail while rotating.  The peak RF position was also altered by reducing the width as much as possible and shifting it to the left of the $I_-$ peak allowing rotation into the ss-RF region rather than away from.  This provided a slight benefit to enhancement during rotation.  The power was also reduced at this position as was very easy to destroy the enhancement gained by irradiating the other absorption line underneath with too much RF.  Figure \ref{no2} shows the results of this type of selective rss-RF.  The total tensor enhancement for this example was 36.5\% (9.7\%).

With ss-RF applied to the two regions indicated during rotation, significant increase to the opposing absorption line could be seen, see Fig. \ref{no2}.  The opposite pedestal grew and the enhanced region moved under the smaller peak.  The opposing peak region also grew in response to the ss-RF.  This is so far the best method for producing a maximum tensor polarization enhancement during continuous DNP and continuous rotation at a constant rate. These results indicate that an absolute tensor polarization enhancement of around 15\% given an initial 50\% vector polarization is viable under the standard conditions of a scattering experiment.  What we conclude from the analysis of the data from rotation is that it is necessary to implement the ss-RF with a power profile that is sensitive to the lineshape as a function the ss-RF frequency, modulation range as well as target rotation rate.  The modulation range should be minimized while using rotation to reduce excess RF-power load to the target.  Rotation can be used to effectively modulate even when the ss-RF is fixed.

There is additional measurement error when rotation is involved due to the time dependence of the spectral diffusion.  This error can likely be reduced with further research and the development of online rss-RF optimization software and measurement tools.  This work is also underway at UVA.

\begin{figure}
\begin{center}
    \includegraphics[width=0.60\textwidth]{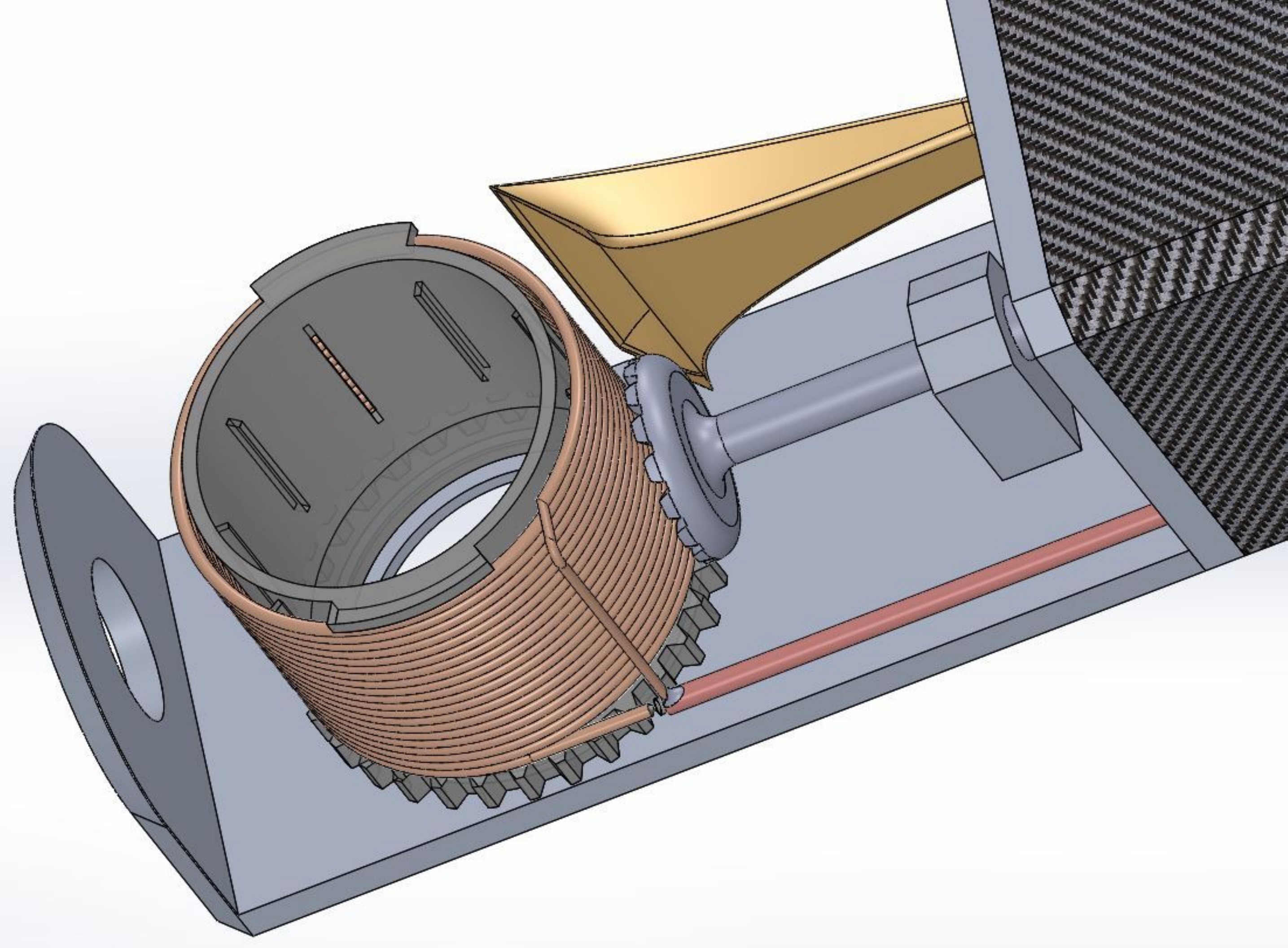}
\end{center}
\caption{Drawing of a possible design for a rss-RF cup that would work in a scattering experiment.  A thin foil lid and bottom would be used directly in the beam path.  The microwave horn and the rotation drive are also visible.}
\label{rcup}
\end{figure}

Figure \ref{rcup} shows a model of a rotating target which would be similar to what would be used in a scattering experiment.  The front and back of the cup would be made of thin foil for the beam to pass through, and the cup would be made to rotate around by a gear driven from the perimeter of the cup.  The ss-RF coil would be fixed to the target ladder and not in contact with the rotating part of the cup, while the NMR coil would be fixed on  the inside of the ss-RF coil.  The cup that was used to take the measurements presented here is similar to this design.

\subsection{Other RF Techniques}

The MAS technique is standard in solid-state NMR frequently used in NMR spectroscopy to suppress dipolar broadening \cite{andrew}.  The local field produced at an observed spin has components with symmetry like those of second rank spherical harmonics. When the sample is spun, local fields orthogonal to the spinning axis are rendered time dependent and modulate the NMR signal at multiples of the spin rate. The resulting NMR signal contains a set of side bands at multiples of the spin rate and a center band whose width is determined by the distribution of local fields along the rotation axis. For dipolar fields, the component along a spinning axis oriented at angle $\beta$ with respect to the holding field is reduced by a factor of $P_2(\text{cos}\beta)$. For MAS, $P_2(\text{cos}\beta)=0$, and dipolar broadening can be greatly suppressed. Complete suppression requires that the spin rate be much larger than the magnitude of the static coupling (usually tens of kHz).  With high frequency spinning, the side band intensities reduce to zero.

Some interactions, such as those from electric quadrupole fields, do not average completely from MAS. In addition, if the rotation is not at the magic angle, the averaging of all the interactions will not be complete.  We expect that there is an avenue of research of high frequency rotation off the magic angle that may also provide additional tensor polarization enhancement for the solid-state spin-1 target.  There are, of course, challenges of high frequency rotation in a DNP environment, let alone in a scattering experiment with continuous beam.

Additional techniques are also being investigated that explore tools like adiabatic fast passage (AFP) of the solid targets of interest.  Much of the concern with using low tensor polarization in a small asymmetry measurement comes from the systematic uncertainty associated with time dependent drifts.  These types of uncertainties can be reduced by either using two targets of different helicity in the beam-line simultaneously or flipping the target polarization more frequently.  In the case of tensor polarization, AFP can also be used to put the target in an enhanced state for a time, as well as to put the target in an unpolarized state for a time.  The flip between these states using AFP is on the order of milliseconds.  Unfortunately, these techniques work best after the DNP has stopped, but it could be used as part of a continuous averaging approach.  It is also possible to maximize tensor enhancement and mitigate vector polarization using RF similar RF techniques.  This allows the separation of some tensor and vector observables and provides a much cleaner extraction.  These and other approaches will be addressed in further publications.

\section{Conclusion}
\label{con}
To achieve the highest figure-of-merit (FOM) in scattering experiments which require a solid-state tensor polarized target, it is necessary to maximize the tensor polarization throughout the beam
target interaction time.  Optimization of the tensor polarization of the spin-1 target ensemble can be achieved by applying ss-RF irradiation at select frequencies with the necessary
power.  Experimental results are provided demonstrating the degree of enhancement using the ss-RF technique to manipulate and measure the CW-NMR signal.  Optimization is achieved using a critical RF decay constant that is controlled by the amplitude of the RF.  The RF partially saturates
the overlapping regions of the Pake doublet corresponding to $P_{zz}(R)<0$ for positive tensor polarization. Enhancement results from reducing the negative contributions to the tensor polarization while simultaneously increasing the positive contributions.

Additional tensor polarization enhancement is achieved by slow rotation of the sample while applying the manipulating RF (rss-RF).  This resultes in a maximized absolute tensor polarization enhancement of around 15\% given an initial 50\% vector polarization.

These measurements are the first results from a set of experimental techniques to enhance tensor polarization developed at the University of Virginia.  This research is still early on in its development and much more research is needed and intended.  Along with new RF technology and techniques, it is necessary to develop software that can optimally enhance and measure the CW-NMR signal in the DNP environment.  Real-time enhancement and measurement technology is critical.  Having the NMR system configured and synchronized to sample quickly between the ss-RF irradiation cycles will also help to extract higher resolution NMR data and maximize the overall FOM of the scattering experiment.

\section*{Acknowledgement}
The authors thank Chris Keith of the Jefferson Lab Target Group for thoroughly reading this manuscript.
We also thank Mikhail Yurov for his contributions to these measurements and the development of the initial software used to perform ss-RF and rss-RF.  His commitment to good scientific practice must be acknowledged.  Figure \ref{cup} and \ref{rcup} were produced by Carlos Ramirez using SolidWorks.
All NMR data presented here was taken at the University of Virginia Solid Polarized Target Lab. This work was supported by DOE contract DE-FG02-96ER40950.

\end{document}